\documentclass{aastex631}
\newcommand{\Cloudy}{{\sc Cloudy}}

\usepackage[graphicx]{realboxes}

\graphicspath{{./}{figures/}}
\accepted{for publication in ApJ}
\begin{document}

\title{X-ray spectroscopy in the microcalorimeter era 4: Optical depth effects on the soft X-rays studied with {\Cloudy}}

\email{priyanka.chakraborty@cfa.harvard.edu}

\author[0000-0002-4469-2518]{P. Chakraborty}
\affiliation{Center for Astrophysics $\vert$  Harvard $\&$ Smithsonian \\
Cambridge, MA, USA}
\affiliation{University of Kentucky \\
Lexington, KY, USA}
\author[0000-0003-4503-6333]{G. J. Ferland}
\affiliation{University of Kentucky \\
Lexington, KY, USA}
\author[0000-0002-8823-0606]{M. Chatzikos}
\affiliation{University of Kentucky \\
Lexington, KY, USA}
\author[0000-0002-9378-4072]{A. C. Fabian}
\affiliation{Institute of Astronomy\\ 
Madingley road, Cambridge CB3 0HA, UK}
\author[0000-0002-4622-4240]{S.Bianchi}
\affiliation{Dipartimento di Matematica e Fisica, Universit\`a degli Studi Roma Tre, via della Vasca Navale 84, I-00146 Roma, Italy \\
}
\author[0000-0002-2915-3612]{F. Guzm\'an}
\affiliation{University of North Georgia \\
Dahlonega, GA, USA}
\author[0000-0002-3886-1258]{Y. Su}
\affiliation{University of Kentucky \\
Lexington, KY, USA}

\begin{abstract}

In this paper, we discuss atomic processes modifying the soft X-ray spectra in
from optical depth effects like photoelectric absorption and electron scattering suppressing 
the soft X-ray lines. 
We also show the enhancement in soft X-ray line intensities in a photoionized environment via continuum pumping.
We quantify the suppression/enhancement by introducing a ``line modification factor ($f_{\rm mod}$)."

If 0 $\leq$ $f_{\rm mod}$ $\leq$ 1, the line is suppressed, which could be the case in both collisionally-ionized and photoionized systems. If $f_{\rm mod}$ $\geq$ 1, the line is enhanced, which occurs in photoionized systems. Hybrid astrophysical sources are also very common, where the environment is partly photoionized and partly collisionally-ionized. Such a system is V1223 Sgr, an intermediate polar binary.
We show the application of our theory by fitting the first-order Chandra MEG spectrum of V1223 Sgr with a  combination of \textsc{Cloudy}-simulated additive cooling-flow and photoionized models. In particular, we account for the excess flux for O~VII, O~VIII, Ne~IX, Ne~X, and Mg~XI lines in the spectrum found in a recent study, which could not be explained with an absorbed cooling-flow model.

\end{abstract}

\section{Introduction}
The strongest emission lines observed in the soft X-ray spectra of astrophysical plasmas are H-like and He-like lines from elements between carbon and sulfur, and  Fe L-shell lines. Such X-ray emitting plasmas can be collisionally-ionized or photoionized. In collisionally-ionized plasma, ionization
or excitation events mainly occur due to electron-ion collisions. In photoionized plasma, ionization is dominated by the interaction 
between photons and ions/atoms. A hybrid of the two types of plasma is also possible, where the cloud is partly collisionally-ionized
and partly photoionized.

Line intensity in observed X-ray spectra can be enhanced or suppressed depending on the following factors.  If photoionized, lines may
be a) enhanced when electrons bound to the ions absorb photons of appropriate wavelength transitioning to higher energy states followed by radiative cascades, also known as continuum pumping, b) suppressed due to optical depth effects.
If collisionally-ionized, only optical depth effects suppress lines in the X-ray spectra. In the soft X-ray regime, the optical depth 
effects leading to line suppression mainly come from photoelectric absorption of line photons or scattering by electrons. Section \ref{sec:linesuppression}
and Section \ref{sec:enhancement} present more detailed discussion.

The effects of continuum pumping on the optically thin line intensities, the so-called ``Case C", was 
introduced by \citet{1938ApJ....88..422B} and later discussed by \citet{1953ApJ...117..399C} and
\citet{1999PASP..111.1524F}. Their study was later extended to describe ``Case D" \citep{2009ApJ...691.1712L, 2016RMxAA..52..419P},  
the optically thick counterpart of Case C. Recently, \citet{2021ApJ...912...26C} 
discussed Case C and Case D in view of the future microcalorimeter missions line XRISM and Athena,
focusing on line emission from H- and He-like iron.

Optical depth effects on soft X-ray spectra have been previously investigated by \citet{1988MNRAS.231.1139S} and \citet{1992MNRAS.254..277H}, 
who calculated the recombination line intensities of H-like carbon, nitrogen, and oxygen assuming ``Case B" \citep{1938ApJ....88...52B}
and discussed the effects of finite optical depth and dust on H-like ions. Such effects have also been observed in soft X-ray emission spectra. For
instance, recent analysis of soft X-ray spectra from Reflection Grating Spectrometer (RGS) onboard XMM-Newton and High Energy Transmission Grating Spectrometer (HETGS) onboard Chandra explored several optically-thick
astrophysical sources like cataclysmic variables, novae, 
ultraluminous X-ray sources, Seyfert galaxies, elliptical galaxies, X-ray binaries, etc \citep{2002MNRAS.334..805R, 2005AIPC..774..361R, 2008ApJ...680..695O,2011AN....332..337P, 2012A&A...539A..34D, 2020MNRAS.497.2352M, 2021A&A...648A.105A}.

As previously mentioned, there have been extensive studies separately discussing the effects of continuum pumping and 
optical depth effects on X-ray line intensities. The current need is to develop a comprehensive theory that will account for
all the atomic processes changing the soft X-ray spectra. 
This paper aims to present a framework for interpreting the soft X-ray spectra from photoionized and collisionally-ionized plasma using the spectral synthesis code \textsc{Cloudy} \citep{2017RMxAA..53..385F}. A cooling-flow model has been introduced to treat the multi-phase collisionally-ionized systems.

We also show the application of our model on V1223  Sgr, a member  of the  cataclysmic  variable  binary star systems called Intermediate Polars. We 
address the lack of emission from several lower atomic number H- and He-like ions, as mentioned in \citet{2021arXiv210705636I}, with  additive cooling-flow and photoionized models. However, our model will be applicable for any optically thick \footnote{Note that, Cloudy generates a warning when the electron scattering optical depth is greater than 5 (refer to cloudy/source/prt\_{comment}.cpp : 2297), corresponding to a hydrogen column density of 7.5 $\times$ 10$^{24}$ cm$^{-2}$. All our simulations are done for N$_{H}$ $\leq$ 7.5 $\times$ 10$^{24}$ cm$^{-2}$.} source with/without a radiation source.

This paper is the fourth of the series ``X-ray spectroscopy in the microcalorimeter era". The first three papers discussed the atomic processes in a collisionally excited and photoionized astrophysical plasma for H- and He-like iron  for  a  wide  range  of  column densities.
\citet{2020ApJ...901...68C} explored  the  effects  of Li-like  iron  on  the line  intensities of four members of the  Fe  XXV  K$\alpha$ complex - x($2^{3}P_{2}\rightarrow1^{1}S$), y($2^{3}P_{1}\rightarrow1^{1}S$), z($2^{3}S\rightarrow1^{1}S$), and w($2^{1}P\rightarrow1^{1}S$) through  Resonant  Auger  Destruction \citep[RAD,][]{1978ApJ...219..292R, 1990ApJ...362...90B, 1996MNRAS.278.1082R, 2005AIPC..774...99L} and line broadening effects by electron scattering.
\citet{2020ApJ...901...69C}  introduced  a  novel  method  of  measuring  column  density  from  Case  A 
to B  (optically  thin  to  thick)  transition  by  comparing  the  observed  spectra  by  Hitomi  with spectra  simulated  with \textsc{Cloudy}. \citet{2021ApJ...912...26C} established  four 
asymptotic  limits-  Cases  A,  B,  C,  and  D  for  describing  the  line  formation processes  in  H- 
and  He-like  iron  emitting  in  the  X-rays. This paper is a continuation of the framework discussed in the first three papers to study the atomic processes affecting the soft X-ray spectrum.

\section{line suppression due to photoelectric and electron opacity} \label{sec:linesuppression}
Photoelectric absorption of line photons or scattering by electrons can significantly 
suppress line intensities. 
Here we present \textsc{Cloudy} calculations of the net emission, demonstrate line suppression due 
to absorption and scattering, and give simple analytical estimates of the physical
processes which suppress the line emission.

\textsc{Cloudy} actually determines the line emission by solving a fully coupled system of equations
giving radiative transitions between levels, as originally described in Section III of 
\citet{1989ApJ...347..640R}.
The numerical results can be understood by considering the following approximations to the line transfer. 

The fraction of photons that survives (which we call  {$f_{\rm mod}$} or the ``line modification factor") after N scatterings that is  subject to photoelectric absorption and electron scattering is determined by the following equation:
\begin{equation}
\label{eqn:frac0}
 {f_{\rm mod}} =  (1-P_{\rm photoelectric}-P_{\rm scattering})^{N},
\end{equation}

where N is the mean number of scatterings experienced by a line photon  with a line-center optical depth $\tau$ before escaping an optically thick cloud \citep{1979ApJ...229..274F}:

\begin{equation}
\label{eqn:N}
\rm{N} =\frac{{{ 1.11\tau ^{1.071}}}}{{1+{\rm{({log\tau/5.5}})^5}}} 
\end{equation}

$P_{\rm photoelectric}$ is the probability of photoelectric absorption per scattering, and is given by the following ratio of opacities:
\begin{equation}
\label{eqn:ESCprob}
{P}_{\rm photoelectric} = \frac{{{ \kappa_{photoelectric}}
}}{{\kappa_{total}}}
\end{equation}
\noindent
$P_{\rm scattering}$ is the probability per scattering that a line photon will be scattered off free electrons due to the very large Doppler shift they receive and be removed from the Doppler core of the line (refer to section 7.1 in \citet{2020ApJ...901...68C} for a detailed discussion):
\begin{equation}
\label{eqn:ESCprob}
{P}_{\rm scattering} = \frac{{{ \kappa_{scattering}}
}}{{\kappa_{total}}}
\end{equation}
\noindent
where the total opacity is given by
\begin{equation}
\kappa_{total}= \kappa_{line}+ \kappa_{photoelectric}+ \kappa_{scattering}
\end{equation}

\begin{figure}[h!!]
\centering
\includegraphics[scale=0.5]{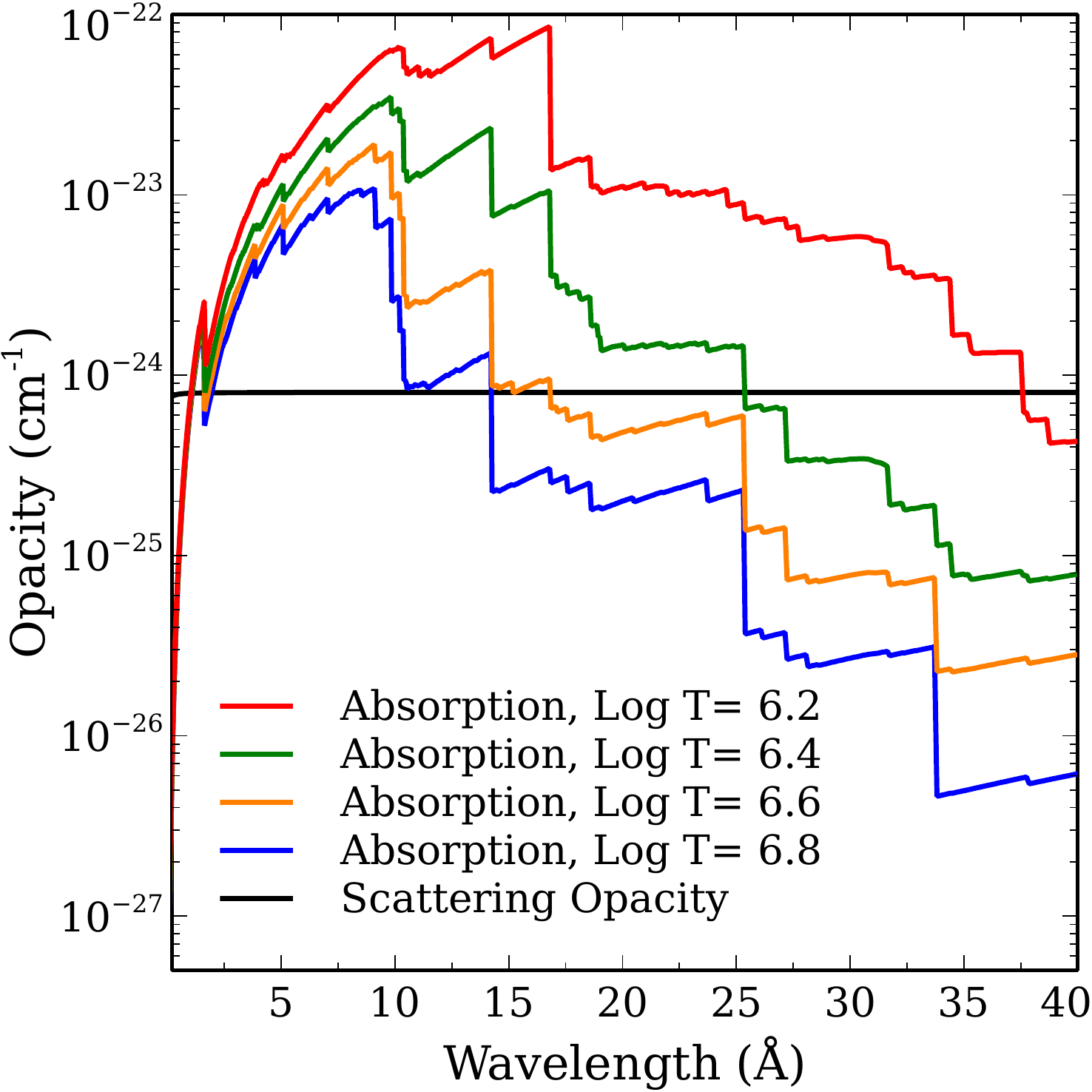}
\caption{Continuum opacities plotted as a function of wavelength for a cloud with N$_{H}$=10$^{23}$ cm$^{-2}$. Red, green, 
yellow and blue solid lines show the absorption opacities (photoelectric) for log T=6.2, 6.4, 6.6, and 6.8, respectively. The black 
solid line shows the electron scattering opacity. The opacity is in units of cm$^{-1}$, the unit of opacity appearing in Kirchhoff-Planck law is 
cm$^{-1}$.}
\label{fig:opacity}
\end{figure}

For the most part,  {$f_{\rm mod}$} gives an estimate for the fraction of photons surviving after photoelectric absorption and electron scattering. 
Figure \ref{fig:opacity} shows photoelectric absorption opacities ($\kappa_{photoelectric}$) for temperatures between log~ T(Kelvin) = 6.2 and log~T(Kelvin) = 6.8 and the electron scattering opacity ($\kappa_{scattering}$), assuming a thermal collisional equilibrium at T.  A hydrogen density (n(H)) of 1 cm$^{-3}$ was assumed, corresponding to an electron density of $\sim$ 1.2 cm$^{-3}$, hydrogen and helium being the two most abundant elements in the Universe. The electron scattering opacity
was evaluated by multiplying the electron scattering cross section (6.65 $\times$ 10$^{-25}$ cm$^{2}$ ) with its density.
Scattering is more important at the longer wavelengths and higher temperatures,
while absorption is more important at the shorter wavelengths and
lower temperatures. Electron scattering opacity has been estimated

Other optical depth effects such as the Case A to B transition and Resonant Auger Destruction can suppress/amplify selected line intensities too \citep{2020ApJ...901...68C,2020ApJ...901...69C}. Numerical \textsc{Cloudy} simulations on optical depth effects shown in this paper include all the atomic processes that can change line intensities. However, we found these effects insignificant compared to photoelectric absorption and electron scattering in the soft X-ray regime. 
The discussion on the optical depth effects on soft X-ray emission continues in Section \ref{sec:softX-ray} for the collisionally-ionized and photoionized cases.

Note that equation \ref{eqn:frac0} applies in  collisionally-ionized environments, where there is no external radiation source.
In irradiated photoionized environments, {$f_{\rm mod}$} will also have contributions from radiative 
cascades prompted by absorption of line photons from the external radiation source, also known as ``continuum pumping". This is further elaborated in Section \ref{sec:enhancement} and Section \ref{photo_emission}.

\begin{figure}[h]
\centering
\includegraphics[scale=0.75]{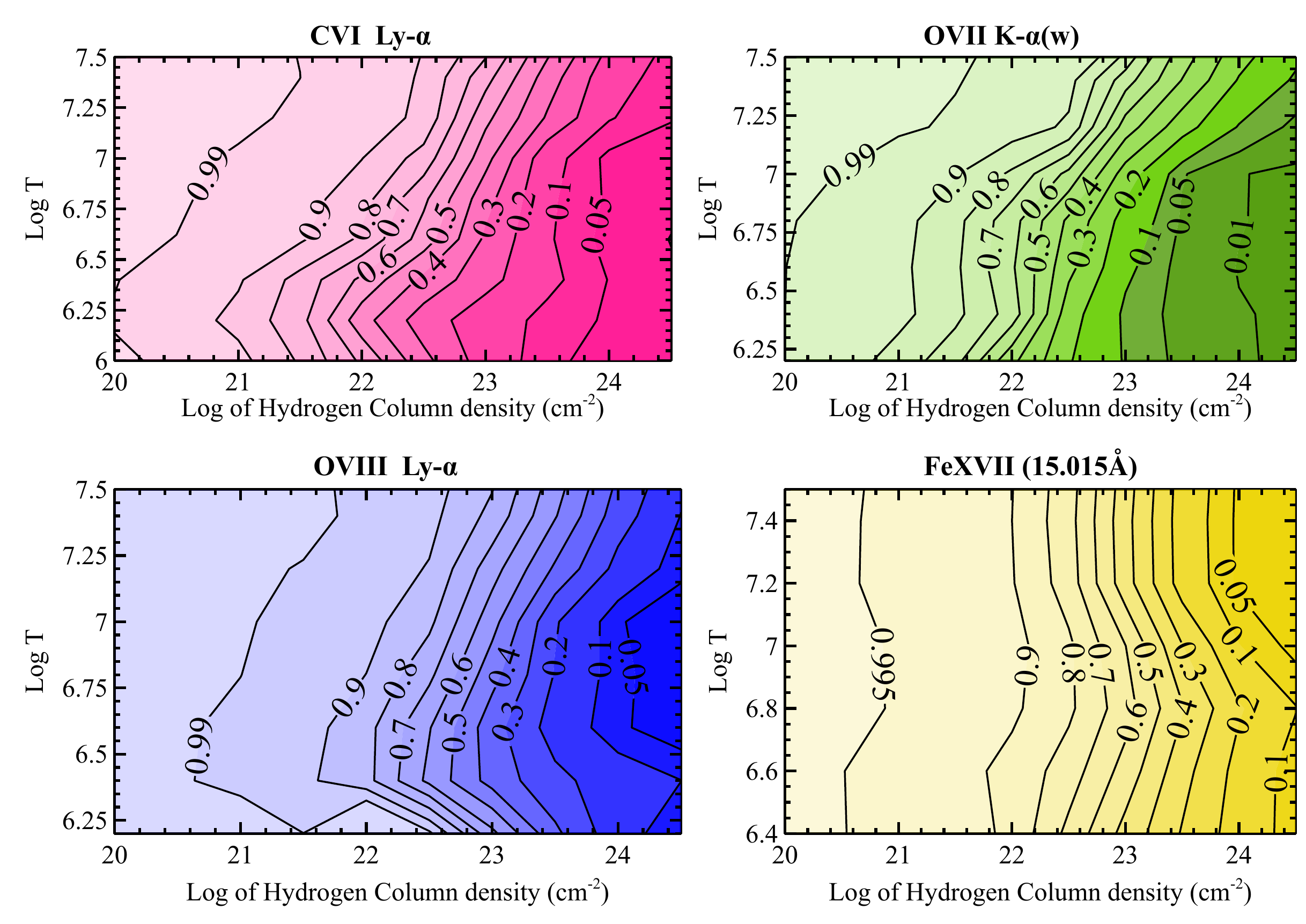}
\caption{Line modification factors  in C~VI Ly-$\alpha$, O~VII K-$\alpha$ (w), O~VIII Ly-$\alpha$, and Fe~XVII, in a collisionally-ionized environment, plotted as a function of column density and temperature. }
\label{fig:supp}
\end{figure}

 \begin{figure}[h]
\centering
\includegraphics[scale=0.75]{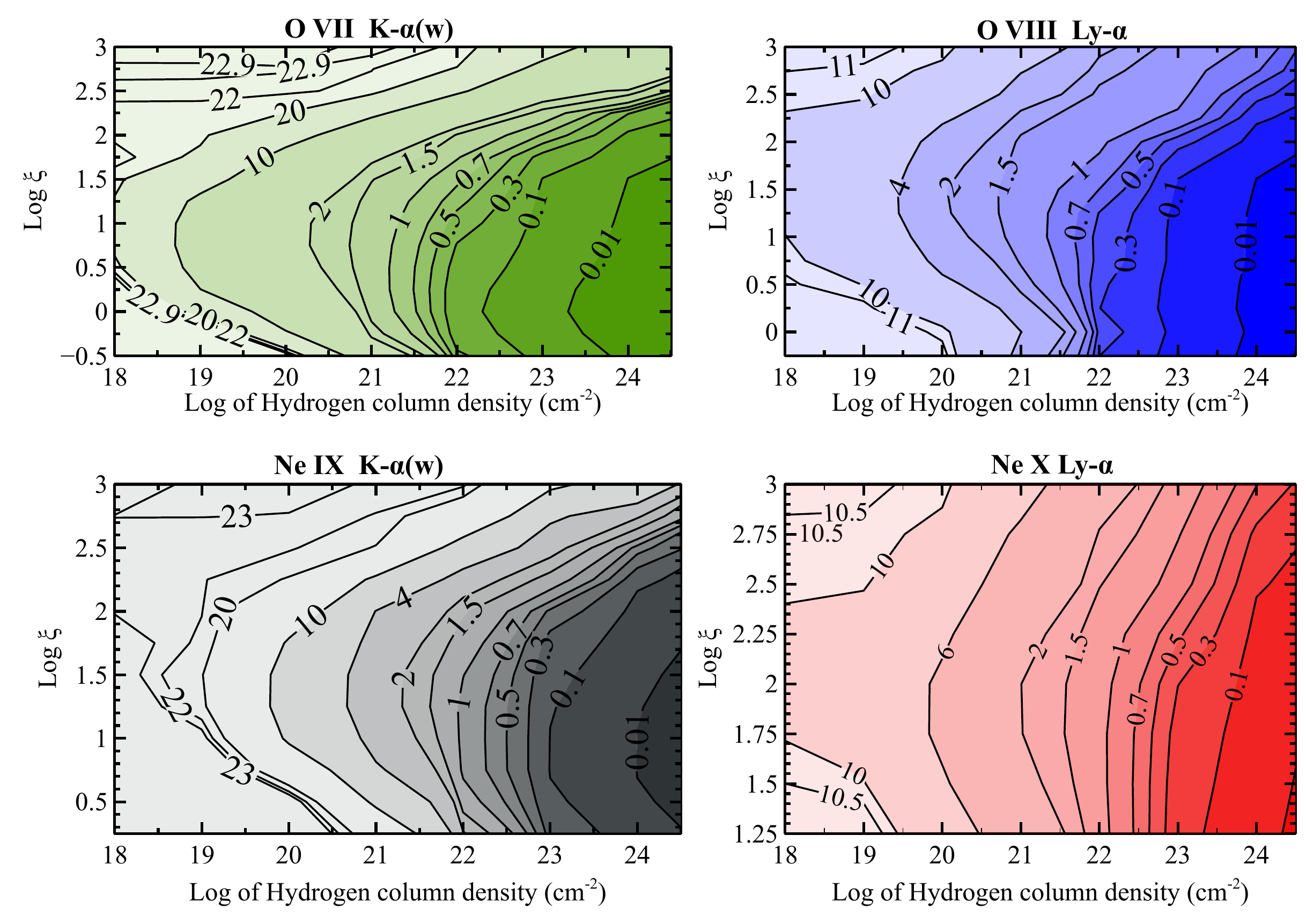}
\caption{Line modification factor  in  O~VII K-$\alpha$ (w), O~VIII Ly-$\alpha$, Ne~IX K-$\alpha$ (w), Ne~X Ly-$\alpha$,  as a function of column density and  ionization parameter ($\xi$).}
\label{f:photoionized}
\end{figure}

\section{line enhancement due to continuum pumping} \label{sec:enhancement}

In photoionized systems, emission lines are enhanced in the presence of external radiation fields due to continuum pumping.
\citet{2021ApJ...912...26C} discussed the concept of continuum pumping and its effects in enhancing the line emission from H- and He-like iron. Lines can be enhanced up to $\sim$ 30 times, and the degree of enhancement decreases with the increase in column density due to the optical depth effects, the so-called Case C to D transition. Refer to section 2 in their paper for the mathematical framework of continuum pumping and figure 1 for a simplified three level representation of Case C and Case D.

This work is an extension of the same concept for the soft X-ray regime. The observed spectra (photoionized) will have contributions from line enhancement due to continuum pumping and line suppression due to photoelectric absorption and electron scattering (further discussed in Section \ref{sec:softX-ray}). Whether the lines are enhanced or suppressed depends on the interplay between these effects.
The enhancement/suppression varies with column density/optical depth, ionization parameter ($\xi$), 
and shape of the incident radiation field. For most of our calculations, we assume a power-law SED:
\begin{equation}
\rm f_{\nu} \propto \nu^{\alpha}
\end{equation}
with $\alpha =-1$, except for Section \ref{obs} where we let $\alpha$ to vary freely.

\section{Optical depth effects on soft X-ray spectra} \label{sec:softX-ray}
\subsection{Collisionally-ionized emission lines}\label{collisional}

The practical way to study the effect of optical depth on line intensities is to compute scaled line intensity ratios of a realistic model that includes all the optical depth effects and a simplistic model excluding optical depth effects. The line ratio stands for what fraction of 
line photons survive.

The realistic model includes absorption and scattering. By default, \textsc{Cloudy} models include absorption but do not include scattering. At X-ray temperatures,
a fraction of the line photons gets largely Doppler-shifted
from their line-center due to scattering off high-speed electrons. The following \textsc{Cloudy} command reports the line intensities excluding these Doppler-shifted photons:

\begin{verbatim}
no scattering intensity 
\end{verbatim}

Refer to Section 6 of \citet{2021ApJ...912...26C} for a detailed discussion on \texttt{no scattering intensity}.

In the simplistic model, absorption was switched off with the command:

\begin{verbatim}
no absorption escape
\end{verbatim}

The smaller the ratio between the realistic and simplistic model, the larger the suppression. This ratio is the numerical equivalent of $f_{\rm mod}$, the line modification factor,  
studied in Section \ref{sec:linesuppression}, which presented an analytical discussion on line suppression in optically thick clouds. Figure \ref{sec:linesuppression} shows this ratio for selected soft X-ray lines evaluated with \textsc{Cloudy}. 

Temperature and column density are physical parameters that can notably change line emission from the collisionally-ionized cloud. Thus, contour plots have been shown where temperature and column density is varied. Metallicity has been fixed at solar for simplicity. In principle, metallicity should not be taken as a free parameter, as optical depth is proportional to metallicity. This section aims to establish the framework using simplified parameters. In Section \ref{obs}, we let metallicity vary as a free parameter along
with other physical parameters to fit the observed spectrum from V1223 Sgr.

Four panels in the figure show the numerical line modification factor for C~VI Ly-$\alpha$ ($\lambda$= 33.736 \AA), OVII K-$\alpha$ 
resonance line (w; $\lambda$=21.602\AA), O~VIII Ly-$\alpha$ ($\lambda$= 18.969 \AA), and Fe~XVII ($\lambda$=15.015\AA)
as a function of the hydrogen column density and temperature for a hydrogen density (n(H)) of 1 cm$^{-3}$. There is significant suppression in carbon and oxygen lines and moderate suppression in the iron lines. For example, for  CVI Ly-$\alpha$ line at the temperature of its peak emissivity,
log T=6.2: at N$_{H}$=10$^{23}$ cm$^{-2}$, the   numerical  line  modification  factor as shown in the top-left panel of the figure is $\sim$ 0.15. This implies, only 
15 \% of the  CVI Ly-$\alpha$ line photons will make it out of the optically thick 
cloud at the mentioned temperature and column density\footnote{We predict the line mean intensity 4$\pi$J averaged over all directions}.

Numerical results in the figure account for the full physics and the variation of the physical conditions across the cloud.
Analytically, equation \ref{eqn:frac0} estimates the fraction of CVI Ly-$\alpha$ photons surviving after
photoelectric absorption and electron scattering, the only two important sources of opacity in the soft X-ray limit.

At the same temperature and column density, $\tau$ $\sim$ 500 (calculated using the \textsc{Cloudy} command \texttt{save line optical depth}), $\kappa_{photoelectric}$= 3.5$\times$ 10$^{-24}$ cm$^{-1}$, 
and $\kappa_{scattering}$= 7.9$\times$ 10$^{-25}$ cm$^{-1}$ (refer to Figure \ref{fig:opacity}). Line opacity ($\kappa_{line}$) can be expressed as a product of  ionic number density ($n_{\rm ion}$),
and absorption cross-section ($\alpha_\nu$) for the corresponding transition:
\begin{equation}\label{e:1}
\kappa_{line} = n_{\rm ion} \alpha_\nu
\end{equation}
Here 
\begin{equation}\label{e:2}
n_{\rm ion} = {\rm carbon} \hspace{1 mm}{\rm abundance} \times {\rm ionic  \hspace{1 mm}fraction  \hspace{1 mm} of}  \hspace{1 mm} \rm C^{5+} \times {\rm n(H)},
\end{equation}
and $\alpha_\nu$ is described by the following equation (see equation 5.23 in hazy2):
\begin{equation}\label{e:3}
\alpha _{\nu(x)}   = 2.24484\times 10^{-14} A_{u,l} \lambda_{\mu m} ^{3}\frac{{g_u}}{g_{l}} \frac{ {\varphi _\nu }(x)}{{u_{\rm Dop}}} 
\end{equation}
A$_{u,l}$ is the transition probability for the transition from level u to level l, $\lambda_{\mu m}$  is the wavelength in micrometer, g$_{u}$ and g$_{l}$ are
statistical weights of levels u and l, u$_{Dop}$ is the Doppler velocity, $\varphi$$_\nu$(x)=1 at line-center.  At logT= 6.2, $u_{\rm Dop}$= $\sqrt{\frac{2kT}{m_{\rm C}}}$ $\sim$ 47 km/s. Using equations \ref{e:1}, \ref{e:2}, and \ref{e:3}, we obtain $\kappa_{line}$= 2.39$\times$ 10$^{-21}$ cm$^{-1}$, which implies $\kappa_{total}$= 2.40$\times$ 10$^{-21}$ cm$^{-1}$, and $f_{\rm mod}$ $\sim$ 0.2. Analytically, 20 \% of the  CVI Ly-$\alpha$ line photons will survive and
make it out of the cloud.

%where $\alpha_\nu$ is inversely proportional to the total Doppler velocity.

Tables \ref{t:c1} and \ref{t:c2} list the  numerical value of $f_{\rm mod}$ at  N$_{H}$ = 10$^{22}$ cm$^{-2}$, N$_{H}$ = 10$^{22.5}$ cm$^{-2}$, and N$_{H}$ = 10$^{23}$ cm$^{-2}$ for log~T = 6.4, and 6.8 for important lines in soft X-ray spectra. The listed $f_{\rm mod}$ values are shown  as example cases  to demonstrate the effect of line modification for a range of astrophysical plasma. The users can generate  $f_{\rm mod}$ for the full parameter space shown in Figure \ref{fig:supp} using the \textsc{Cloudy} command:
\begin{verbatim}
save line intensity 
\end{verbatim}  
along with the \textsc{Cloudy} commands described earlier in Section \ref{collisional}.
$f_{\rm mod}$ gives the fraction of line photons that will be observed by high-resolution telescopes. Note that the emission lines from the lower-Z elements exhibit maximum emissivity at the lower temperatures, where the photoelectric absorption is significant (see Figure \ref{fig:opacity}). This will lead to  a larger suppression in the lines from lower-Z elements than those of the higher-Z elements.

\begin{table*}
\caption{Line modification factor ($f_{mod}$) in collisionally ionized cloud at N$_{H}$ = 10$^{22}$ cm$^{-2}$, N$_{H}$ = 10$^{22.5}$ cm$^{-2}$, and N$_{H}$ = 10$^{23}$ cm$^{-2}$ and log T = 6.4 for important lines in soft X-ray spectra. Lines from lower atomic number elements 
are more suppressed than those of higher atomic number elements , as mentioned in Section \ref{collisional}.}%A ``-" under the f$_{modification}$ column means, the line is too faint to see at that set of parameters.}
\begin{tabular}{lllrrr}
\hline
\multicolumn{5}{r}{$f_{\rm mod}$} \\
\cline{4 -6}
 Transitions  &$\lambda$ (\AA) & label  & N$_{H}$=10$^{22}$ cm$^{-2}$ & N$_{H}$=10$^{22.5}$ cm$^{-2}$ & N$_{H}$=10$^{23}$ cm$^{-2}$    \\
\hline
C VI & 33.736  & Ly$\alpha$& 0.57& 0.43 & 0.20\\
N VI  & 29.082& K$\alpha$ (x) &0.40 & 0.21  & 0.05\\
N VI  & 28.787  &  K$\alpha$ (w) & 0.37 & 0.18 & 0.04\\
N VII  & 24.781  &  Ly$\alpha$& 0.83& 0.69  & 0.24 \\
O VII & 22.101  & K$\alpha$ (z)& 0.73 & 0.35 & 0.13\\
O VII & 21.807 &  K$\alpha$ (y) & 0.71& 0.33 & 0.12\\
O VII & 21.804  & K$\alpha$ (x)& 0.68& 0.31& 0.11\\
O VII & 21.602 & K$\alpha$ (w) & 0.61& 0.24 & 0.08\\
O VIII & 18.969 & Ly$\alpha$& 0.82 & 0.58 & 0.35\\
Fe XVII & 15.262 & 3D$^\dagger$ & 0.84 & 0.61 & 0.40\\
Fe XVII & 15.015 & 3C$^\dagger$ &0.86 & 0.62& 0.46\\
Ne IX & 13.699  &  K$\alpha$ (z) & 0.99  & 0.97& 0.65\\
Ne IX & 13.553 &  K$\alpha$ (y) & 0.99& 0.95&  0.61\\
Ne IX & 13.550  & K$\alpha$ (x) & 0.96 & 0.87 & 0.53\\
Ne IX & 13.447 &  K$\alpha$ (w) & 0.81& 0.64 & 0.33\\
%S XVI & 4.729 & Ly$\alpha$ &- &-& -\\ 
%S XV  & 5.039 & K$\alpha$ (w) &-  &-& -  \\
%S XV  & 5.063 &  K$\alpha$ (x) &- &-&  -   \\
%S XV  & 5.066 &  K$\alpha$ (y) & -   &-& - \\
%%S XV  & 5.102 & K$\alpha$ (z) & -&-&- \\
%Si XIV  & 6.182 &   Ly$\alpha$  &-  &-&-\\
%Si XIII  & 6.648 &   K$\alpha$ (w) &-  &-& -\\
%Si XIII & 6.685 &   K$\alpha$ (x) & - &-&-\\
%Si XIII & 6.688 &  K$\alpha$ (y) & - &-& - \\
%Si XIII  & 6.740 &  K$\alpha$ (z)& - &-&-\\
%Mg XII & 8.421& Ly$\alpha$ & -&-& -\\
%Mg XI & 9.169& K$\alpha$ (w) &-  &-& -\\
%Mg XI & 9.228  & K$\alpha$ (x) &-  &-& -\\
%Mg XI & 9.231  & K$\alpha$ (y) & -&-& -\\
%Mg XI & 9.314  & K$\alpha$ (z)& - &-& -\\
%Ne X & 12.134 & Ly$\alpha$&-  &-& -\\\
%Fe XVII & 17.051 & -& -& -& -\\
%Fe XVII & 17.096 & -&- & - & -\\

\hline

\footnote{$^\dagger$ Labeled by Brown et al. (1998)}
\end{tabular}
\label{t:c1}
\end{table*}

\begin{table*}
\caption{Line modification factor ($f_{\rm mod}$) in collisionally ionized cloud at N$_{H}$ = 10$^{22}$ cm$^{-2}$, N$_{H}$ = 10$^{22.5}$ cm$^{-2}$, and N$_{H}$ = 10$^{23}$ cm$^{-2}$ and log T = 6.8 for important lines in soft X-ray spectra.
Suppression is less than that listed in Table \ref{t:c1} as photoelectric absorption decreases with an increase in temperature.
Lines from lower atomic number elements are more suppressed.} %A ``-" under the f$_{modification}$ column means, the line is too faint to see at that set of parameters.}
\begin{tabular}{lllrrr}
\hline
\multicolumn{5}{r}{$f_{\rm mod}$} \\
\cline{4 -6}
 Transitions  &$\lambda$ (\AA) & label  & N$_{H}$=10$^{22}$ cm$^{-2}$ & N$_{H}$=10$^{22.5}$ cm$^{-2}$ & N$_{H}$=10$^{23}$ cm$^{-2}$    \\
\hline
%S XVI & 4.729 & Ly$\alpha$ &- &-& -\\ 
O VIII & 18.969 & Ly$\alpha$& 0.89 & 0.72 & 0.43\\
Fe XVII & 17.096 & M2$^\dagger$ &1.0& 0.99 & 0.97\\
Fe XVII & 17.051 &  3G$^\dagger$ & 1.0& 1.0& 1.0\\
Fe XVII & 15.262 & 3D$^\dagger$ & 0.88 & 0.71 & 0.50\\
Fe XVII & 15.015 & 3C$^\dagger$ &0.91 & 0.81& 0.64\\
Ne IX & 13.699  &  K$\alpha$ (z) & 0.95  & 0.88 & 0.74\\
Ne IX & 13.553 &  K$\alpha$ (y) & 0.95 & 0.87 &  0.72\\
Ne IX & 13.550  & K$\alpha$ (x) & 0.94 & 0.85 & 0.69\\
Ne IX & 13.447 &  K$\alpha$ (w) & 0.93& 0.82 & 0.62\\
Ne X & 12.134 & Ly$\alpha$& 0.95  & 0.86& 0.69\\
Mg XI & 9.314  & K$\alpha$ (z)& 0.99 & 0.97& 0.92\\
Mg XI & 9.231  & K$\alpha$ (y) & 0.99& 0.96& 0.90\\
Mg XI & 9.228  & K$\alpha$ (x) &0.97  & 0.93& 0.84\\
Mg XI & 9.169& K$\alpha$ (w) &0.95  & 0.85& 0.71\\
Mg XII & 8.421& Ly$\alpha$ & 1.0 & 1.0 & 0.98\\
Si XIII  & 6.740 &  K$\alpha$ (z)& 1.0 & 1.0 & 1.0\\
Si XIII & 6.688 &  K$\alpha$ (y) & 1.0 & 1.0 & 1.0 \\
Si XIII & 6.685 &   K$\alpha$ (x) & 1.0 & 1.0 & 1.0\\
Si XIII  & 6.648 &   K$\alpha$ (w) &0.97  & 0.90 & 0.80\\
S XV  & 5.102 & K$\alpha$ (z) & 1.0 & 1.0 & 1.0 \\
S XV  & 5.066 &  K$\alpha$ (y) & 1.0   & 1.0 & 1.0 \\
S XV  & 5.063 &  K$\alpha$ (x) & 1.0 & 1.0 &  1.0   \\
S XV  & 5.039 & K$\alpha$ (w) &1.0  & 1.0 & 0.96  \\

%Si XIV  & 6.182 &   Ly$\alpha$  &-  &-&-\\

%O VII & 21.602 & K$\alpha$ (w) & 0.61& 0.24 & 0.08\\
%O VII & 21.804  & K$\alpha$ (x)& 0.68& 0.31& 0.11\\
%O VII & 21.807 &  K$\alpha$ (y) & 0.71& 0.33 & 0.12\\
%O VII & 22.101  & K$\alpha$ (z)& 0.73 & 0.35 & 0.13\\
%N VII  & 24.781  &  Ly$\alpha$& 0.83& 0.69  & 0.24 \\
%N VI  & 28.787  &  K$\alpha$ (w) & 0.37 & 0.18 & 0.04\\
%N VI  & 29.082& K$\alpha$ (x) &0.40 & 0.21  & 0.05\\
%C VI & 33.736  & Ly$\alpha$& 0.57& 0.34 & 0.15\\
\hline

\footnote{$^\dagger$ Labeled by Brown et al. (1998)}
\end{tabular}

\label{t:c2}

\end{table*}

\subsection{Photoionized emission lines}\label{photo_emission}

Photoionized emission lines can be enhanced or suppressed, which is determined by the following two factors.
-a) All Lyman like lines going to the ground state\footnote{Originally Baker et al. (1938) discussed the enhancement in the line emission from atomic hydrogen. Here we have studied emissions from both H- and He-like ions. All the Lyman resonance lines become significantly enhanced, forbidden lines are slightly enhanced.} are enhanced by induced radiative excitation 
of the atoms/ions by continuum photons in the SED. 
-b) Photoelectric absorption and electron scattering in line photons can suppress the line intensities.

The line ratio to be studied here is the scaled ratio of a realistic photoionized model that includes all the optical depth effects and continuum pumping,
and a simplistic model excluding optical depth effects and continuum pumping. 

Section \ref{collisional} listed the {\Cloudy} commands for switching on/off scattering and absorption processes. The continuum
pumping in the simplistic model was switched off using the {\Cloudy} command:

\begin{verbatim}
no induced processes
\end{verbatim}

This ratio between the realistic and simplistic model can be smaller or larger than 1 depending on whether the 
suppression due to optical thickness or the enhancement due to continuum pumping is dominating. 
Therefore, $f_{\rm mod}$ can be smaller or larger than 1, unlike the collisionally-ionized case, where $f_{\rm mod}$ can only be smaller than 1.

 As mentioned in Section \ref{sec:enhancement}, we assumed a power-law SED
with $\alpha$ =-1 for demonstration. Like the previous subsection, we adopt a solar metallicity. 
However, $\alpha$  has been considered a free parameter in  Section \ref{obs} along with variable metallicity to fit the 
HETG spectrum of V1223 Sgr.  Log of hydrogen column density and ionization parameter have been varied to make contour plots in Figure \ref{f:photoionized}. 

Four contour plots in the figure show the numerical modification factor
for  O~VII K-$\alpha$ resonance (w) line (21.602 \AA), O~VIII Ly-$\alpha$ (18.969 \AA), Ne~IX K-$\alpha$ resonance (w) line (13.447 \AA), and  Ne~X Ly-$\alpha$ (12.134 \AA). 
The modification factor can be as large as $\sim$ 20 (enhanced about 20 times), or as small as $\sim$ 0.01 (suppressed about 100 times) 
for the OVII K-$\alpha$ (w) line depending on the values of N$_{H}$ and $\xi$. The OVIII Ly-$\alpha$ and Ne IX K-$\alpha$  emission lines
can be enhanced $\sim$ 10 times, and suppressed $\sim$ 100 times. The Ne X Ly-$\alpha$ line can be enhanced $\sim$ 4 times, and suppressed $\sim$ 100 times.
The line modification factor for important soft X-ray lines emitted from a photoionized cloud is listed in Tables \ref{t:p1} and \ref{t:p2} 
for N$_{H}$ = 10$^{21}$ cm$^{-2}$, N$_{H}$ = 10$^{22}$ cm$^{-2}$, and N$_{H}$ = 10$^{23}$ cm$^{-2}$ for log $\xi$ = 1 and 2.5.  The $f_{\rm mod}$ values for the full parameter space shown in Figure \ref{f:photoionized} can be estimated using the \textsc{Cloudy} command: 
\begin{verbatim}
save line intensity 
\end{verbatim} 

along with previously listed \textsc{Cloudy} commands in this section.

\begin{table*}
\caption{Line modification factor ($f_{mod}$) in a photoionized cloud at N$_{H}$ = 10$^{21}$ cm$^{-2}$, N$_{H}$ = 10$^{22}$ cm$^{-2}$, and N$_{H}$ = 10$^{23}$ cm$^{-2}$ for log $\xi$ = 1 for important lines in soft X-ray spectra.} %A ``-" under the f$_{modification}$ column means, the line is too faint to see at that set of parameters.}
\begin{tabular}{lllrrr}
\hline
\multicolumn{5}{r}{$f_{\rm mod}$} \\
\cline{4 -6}
 Transitions  &$\lambda$ (\AA) & label  & N$_{H}$=10$^{21}$ cm$^{-2}$ & N$_{H}$=10$^{22}$ cm$^{-2}$ & N$_{H}$=10$^{23}$ cm$^{-2}$ \\
\hline
%S XVI & 4.729 & Ly$\alpha$ &- &-& -\\ 
C VI & 33.736  & Ly$\alpha$& 1.05& 0.38 & 0.03\\
N VI  & 29.082& K$\alpha$ (x) & 1.10 &  0.99 & 0.13\\
N VI  & 28.787  &  K$\alpha$ (w) & 1.75 & 0.72 & 0.07\\
N VII  & 24.781  &  Ly$\alpha$& 1.51 & 0.63  & 0.06 \\
O VII & 22.101  & K$\alpha$ (z)& 1.08 & 0.87 & 0.11\\
O VII & 21.807 &  K$\alpha$ (y) & 1.05& 0.67 & 0.07\\
O VII & 21.804  & K$\alpha$ (x)& 1.07& 0.68 & 0.07\\
O VII & 21.602 & K$\alpha$ (w) & 1.29 & 0.53 & 0.08\\
O VIII & 18.969 & Ly$\alpha$& 1.45 & 0.61 & 0.09\\
Ne IX & 13.699  &  K$\alpha$ (z) & 1.15  & 0.72 & 0.08\\
Ne IX & 13.553 &  K$\alpha$ (y) & 1.16 & 0.73 &  0.08\\
Ne IX & 13.550  & K$\alpha$ (x) & 1.15 & 0.67 & 0.07\\
Ne IX & 13.447 &  K$\alpha$ (w) & 3.55 & 0.94 & 0.10\\
Ne X & 12.134 & Ly$\alpha$& 5.20  & 1.28 & 0.13\\
Mg XI & 9.314  & K$\alpha$ (z)& 1.25 & 0.65 & 0.07\\
Mg XI & 9.231  & K$\alpha$ (y) & 1.30& 0.66 & 0.07\\
Mg XI & 9.228  & K$\alpha$ (x) &1.25  & 0.65& 0.07\\
Mg XI & 9.169& K$\alpha$ (w) & 18.84  & 4.24 & 0.43\\
Mg XII & 8.421& Ly$\alpha$ & 10.96  & 4.69 & 0.50\\
Si XIII  & 6.740 &  K$\alpha$ (z)& 1.39 & 0.56 & 0.05\\
Si XIII & 6.688 &  K$\alpha$ (y) & 1.49 & 0.60 & 0.06 \\
Si XIII & 6.685 &   K$\alpha$ (x) & 1.37 & 0.55 & 0.05\\
Si XIII  & 6.648 &   K$\alpha$ (w) &31.60  & 12.11 & 1.21\\
%S XV  & 5.102 & K$\alpha$ (z) & 1.0 & 1.0 & 1.0 \\
%S XV  & 5.066 &  K$\alpha$ (y) & 1.0   & 1.0 & 1.0 \\
%S XV  & 5.063 &  K$\alpha$ (x) & 1.0 & 1.0 &  1.0   \\
%S XV  & 5.039 & K$\alpha$ (w) &1.0  & 1.0 & 0.96  \\

%Si XIV  & 6.182 &   Ly$\alpha$  &-  &-&-\\

%O VII & 21.602 & K$\alpha$ (w) & 0.61& 0.24 & 0.08\\
%O VII & 21.804  & K$\alpha$ (x)& 0.68& 0.31& 0.11\\
%O VII & 21.807 &  K$\alpha$ (y) & 0.71& 0.33 & 0.12\\
%O VII & 22.101  & K$\alpha$ (z)& 0.73 & 0.35 & 0.13\\
%N VII  & 24.781  &  Ly$\alpha$& 0.83& 0.69  & 0.24 \\
%N VI  & 28.787  &  K$\alpha$ (w) & 0.37 & 0.18 & 0.04\\
%N VI  & 29.082& K$\alpha$ (x) &0.40 & 0.21  & 0.05\\
%C VI & 33.736  & Ly$\alpha$& 0.57& 0.34 & 0.15\\
\hline

\end{tabular}
\label{t:p1}
\end{table*}

\begin{table*}
\caption{Line modification factor ($f_{\rm mod}$) in a photoionized cloud at $N_{\rm H}$ = 10$^{21}$ cm$^{-2}$, $N_{\rm H}$ = 10$^{22}$ cm$^{-2}$, and $N_{\rm H}$ = 10$^{23}$ cm$^{-2}$ for log $\xi$ = 2.5 for important lines in soft X-ray spectra.} %A ``-" under the f$_{modification}$ column means, the line is too faint to see at that set of parameters.}
\begin{tabular}{lllrrr}
\hline
\multicolumn{5}{r}{$f_{\rm mod}$} \\
\cline{4 -6}
 Transitions  &$\lambda$ (\AA) & label  & N$_{H}$=10$^{21}$ cm$^{-2}$ & N$_{H}$=10$^{22}$ cm$^{-2}$ & N$_{H}$=10$^{23}$ cm$^{-2}$    \\
\hline
%S XVI & 4.729 & Ly$\alpha$ &- &-& -\\ 
C VI & 33.736  & Ly$\alpha$& 7.27& 2.43 & 1.27\\
N VI  & 29.082 & K$\alpha$ (x) &0.99 & 0.96  & 0.95\\
N VI  & 28.787  &  K$\alpha$ (w) & 23.74 & 21.32 & 13.37\\
N VII  & 24.781  &  Ly$\alpha$& 7.86 & 2.73  & 1.24 \\
O VII & 22.101  & K$\alpha$ (z)& 0.99& 0.96 & 0.90\\
O VII & 21.807 &  K$\alpha$ (y) & 1.0 & 0.98 & 0.91\\
O VII & 21.804  & K$\alpha$ (x)& 0.99& 0.97& 0.91\\
O VII & 21.602 & K$\alpha$ (w) & 22.06 & 11.12 & 6.74\\
O VIII & 18.969 & Ly$\alpha$& 3.65 & 1.59 & 1.39\\
Ne IX & 13.699  &  K$\alpha$ (z) & 1.26  & 1.03 & 1.01\\
Ne IX & 13.553 &  K$\alpha$ (y) & 1.29 & 1.04 &  1.02\\
Ne IX & 13.550  & K$\alpha$ (x) & 1.27 & 1.03 & 1.01\\
Ne IX & 13.447 &  K$\alpha$ (w) & 19.85 & 11.88 & 4.33\\
Ne X & 12.134 & Ly$\alpha$& 3.55  & 2.60 & 2.31\\
Mg XI & 9.314  & K$\alpha$ (z)& 1.16 & 1.04& 1.03\\
Mg XI & 9.231  & K$\alpha$ (y) & 1.20& 1.08 & 1.07\\
Mg XI & 9.228  & K$\alpha$ (x) &1.16  & 1.04 & 1.03 \\
Mg XI & 9.169& K$\alpha$ (w) & 16.75  & 5.02 & 1.91\\
Mg XII & 8.421& Ly$\alpha$ & 4.88 & 1.87 & 0.95\\
Si XIII  & 6.740 &  K$\alpha$ (z)& 1.10 & 1.08 & 1.07\\
Si XIII & 6.688 &  K$\alpha$ (y) & 1.16 & 1.15 & 1.14 \\
Si XIII & 6.685 &   K$\alpha$ (x) & 1.09 & 1.08 & 1.06\\
Si XIII  & 6.648 &   K$\alpha$ (w) &11.83  & 3.54 & 1.38\\
Si XIV &  6.182& Ly$\alpha$  & 5.24& 2.00 &0.99\\ 
%S XV  & 5.102 & K$\alpha$ (z) & 1.15 & 1.13 & 1.11 \\
%S XV  & 5.066 &  K$\alpha$ (y) & 1.30   & 1.29 & 1.15 \\
%S XV  & 5.063 &  K$\alpha$ (x) & 1.10 & 1.09 &  1.04   \\
%S XV  & 5.039 & K$\alpha$ (w) &14.35  & 4.35 & 1.64  \\
%S XVI &  6.182& Ly$\alpha$  & 8.24& 3.22 & 1.31\\ 
%Si XIV  & 6.182 &   Ly$\alpha$  &-  &-&-\\
%O VII & 21.602 & K$\alpha$ (w) & 0.61& 0.24 & 0.08\\
%O VII & 21.804  & K$\alpha$ (x)& 0.68& 0.31& 0.11\\
%O VII & 21.807 &  K$\alpha$ (y) & 0.71& 0.33 & 0.12\\
%O VII & 22.101  & K$\alpha$ (z)& 0.73 & 0.35 & 0.13\\
%N VII  & 24.781  &  Ly$\alpha$& 0.83& 0.69  & 0.24 \\
%N VI  & 28.787  &  K$\alpha$ (w) & 0.37 & 0.18 & 0.04\\
%N VI  & 29.082& K$\alpha$ (x) &0.40 & 0.21  & 0.05\\
%C VI & 33.736  & Ly$\alpha$& 0.57& 0.34 & 0.15\\
\hline
\end{tabular}
\label{t:p2}
\end{table*}

\begin{table}
\centering{
\caption{\label{t:o1}Archival Chandra ACIS-S/HETG observation of V1223 Sgr.
  }}
\begin{tabular}{ccccc}
\hline
Observation ID  & Start Date/Time  &  Exposure(ksec)\\
\hline
649 & 2000-04-30/16:19:51 & 51.48 \\
\hline
\end{tabular}
\end{table}

\section{Application on V1223 Sgr}\label{obs}
This section shows the application of the theory discussed in Sections \ref{sec:linesuppression}, \ref{sec:enhancement}, and \ref{sec:softX-ray} with a slightly 
modified approach to treat the multi-phase plasma for the collisionally ionized case (refer to Section \ref{multi-temp}).
We show the application on  the Intermediate Polar V1223 Sgr, consisting of a white dwarf accreting material from a low mass secondary star.
The matter stripped from the secondary star is magnetically channeled onto a truncated accretion disk around the white dwarf \citep{1994PASP..106..209P}.  
The X-rays from intermediate polars are generated when the infalling  high-velocity (3000 - 10,000 km/s) matter encounters a shock while falling onto the white dwarf surface. The infalling particles then decelerate
further in a subsonic cooling column before hitting the surface of the white dwarf. The inner flow produces strong X-ray 
emission.

As displayed in Figs \ref{fig:supp} and \ref{f:photoionized}, the optical depth effects are expected to modify the important lines in the soft X-ray spectra at high column densities (N$_{H}$ $\geq$ 10$^{22.5}$ cm$^{-2}$). V1223 Sgr is an ideal candidate for applying our theory, as \citet{2021arXiv210705636I} estimated that  this intermediate polar has a fairly large column density  ($\sim$ 10$^{23}$ cm$^{-2}$). 
They used a combination of a cooling flow model (containing emission lines from multi-temperature collisionally-ionized plasma, and bremsstrahlung), an ionized complex absorber model, and  X-ray reflection to model the summed first-order Chandra MEG spectrum of V1223 Sgr. They reported an excess flux from O~VII, O~VIII,  Ne~IX,  Ne~X, and Mg~XI lines (see figure 3 in their paper) which could not be described with their model, and implied that these lines could have a photoionization origin. The photoionized emission in CV's likely generates in the pre-shock region that is ionized by the X-ray radiation emitted in the post-shock region  \citep{2010ApJ...711.1333L}. We have also fitted the same spectrum with {\Cloudy} and the detailed 
 models are described in Sections \ref{multi-temp} and \ref{Cloudy model}.
 
 \subsection{Data reduction}\label{data reduction}
We redo the Chandra/HETG data analysis of V1223 Sgr, previously done by  \citet{2021arXiv210705636I}.
The observational log of the spectrum is listed in Table \ref{t:o1}.
We obtained the first-order MEG spectrum with the corresponding response files from the  TGCat archive \citep{2011AJ....141..129H, 2013ascl.soft03012H}.
Data processing was performed using {\tt CIAO - 4.13} \citep{2006SPIE.6270E..1VF} and {\tt CALDB} version 4.9.4.
All first-order spectra and associated responses for MEG were combined using the CIAO tool {\tt combine\_spectra}. 
The combined spectrum was grouped using the CIAO tool \texttt{dmgroup} with at least 30 counts per bin. 
This particular lower limit has been set to maintain sufficient counts for the spectral fitting while preserving the resolution of the spectra.
Black data points in Figure \ref{fig:v1223} represent the final spectrum \footnote{
The orbital period of V1223 Sgr is 3.366 hrs \citep{2004A&A...419..291B}. The source orbits about
4 times during the 51.48 ksec observation time. . The spectrum is therefore accumulative over 4 orbits.
}

\subsection{Modeling multi-phase plasma with cooling-flow \textsc{Cloudy} simulation}\label{multi-temp}

The atomic physics and the line modification factor described for collisionally-ionized plasma in Section \ref{collisional} was done for a single-temperature static model for the purpose of demonstration. V1223 Sgr is a multi-phase system, which requires the consideration of the evolution of plasma cooling with time. \citet{2015MNRAS.446.1234C}
used {\Cloudy} to compute  the spectrum of a unit volume of gas  in collisional ionization equilibrium cooling from $\sim$ 10$^{8}$ K to  10$^{4}$ K. We adopt a similar approach with a finite volume of gas and variable column density. 
The cooling-flow model used in our {\Cloudy} simulation consists of a multi-phase plasma that cools from 4$\times$10$^{8}$ K.
The upper limit in temperature has been set following the previous studies with intermediate polars \citep{2008A&A...489.1243A, 2021arXiv210705636I}.
We attempted to let the gas cool down to $10^{4}$ K, but the solution was numerically unstable below 2 $\times$10$^{6}$ K, so we presented the result  down to that temperature. The impact of neglecting the emission below 2 $\times$ 10$^{6}$ K has been discussed in Section \ref{Cloudy model}.
The observed spectrum was well fitted by the cooling-flow model except for the emission from O~VII, O~VIII,  Ne~IX,  Ne~X, and Mg~XI, which is in agreement with \citet{2021arXiv210705636I}. We attempted to fit the excess emission from these lines with a photoionized model using the diffuse emission from the cooling-flow as the SED to photoionize the gas in the pre-shock region. The cooling-flow SED was extracted with the \textsc{Cloudy} command \texttt{save continuum}. The following subsection further describes the parameters and grids employed in our cooling-flow simulation.

\subsection{\textsc{Cloudy} models fitting the observed spectrum}\label{Cloudy model}

We fit the observed spectrum with: (a) pure cooling-flow model; (b) cooling-flow + photoionized model;  (c) pure photoionized model; and (d) cooling-flow + photoionized model without considering the optical depth and continuum pumping effects discussed in Sections \ref{sec:linesuppression}, \ref{sec:enhancement}, and \ref{sec:softX-ray}.  The source of photoionizing radiation in the last three models is the diffuse radiation emitted by the cooling-flow plasma. The \textsc{Cloudy} commands to switch off optical depth effects and continuum pumping are:\\
\texttt{ no absorption escape}\\
\texttt{ no induced processes} 

%To account for the excess flux addressed in \citet{2021arXiv210705636I}, we add a photoionized model with a cooling-flow model simulated with \textsc{Cloudy}. 
We use the \textsc{Cloudy}/XSPEC interface \citep{2006PASP..118..920P} to import the models into XSPEC \citep{1996ASPC..101...17A}. Galactic absorption was included using the XSPEC routine {\tt tbabs}, with the absorbing hydrogen column density\footnote{https://heasarc.gsfc.nasa.gov/cgi-bin/Tools/w3nh/w3nh.pl} set to 0.09$\times$10$^{21}$ cm$^{-2}$ towards V1223 Sgr.
 The complete models as imported to XSPEC are: (a) {\tt tbabs}* {\tt atable}{\tt (cooling-flow.fits)} (b) {\tt tbabs}*
({\tt atable}{\tt (cooling-flow.fits)} + {\tt atable}{\tt (photoionized.fits)}) (c) {\tt tbabs}* {\tt atable}{\tt (photoionized.fits)} (d) {\tt tbabs}*({\tt atable}{\tt (cooling-flow.fits)} + {\tt atable}{\tt (photoionized.fits)}) enabling  \textsc{Cloudy} commands stated in the previous paragraph. 

A chi-square statistics was used to fit the spectrum. This is a straightforward approach for fitting the spectrum while considering all the atomic processes happening in the cloud, including absorption and scattering. The cooling-flow model has grids on log of column density (N$_{H_{coll}}$), and metallicity. The photoionized model has grids on log of column density (N$_{H_{photo}}$), log of ionization parameter ($\xi$),  and metallicity. For b) and d), we tie the metallicity of the cooling-flow and photoionized models, and let log~N$_{H_{coll}}$, log N$_{H_{photo}}$, log $\xi$, and normalization of the models (N$_{photo}$, N$_{coll}$) vary freely. A solar abundance table by \citet{2009ARA&A..47..481A} was used for all our calculations. 
An electron density of 10$^{14}$ cm$^{-3}$ was assumed \citep{2004A&A...419..291B, 2006csxs.book..421K}\footnote{ \citet{2004ApJ...610..991H} showed that line intensities/ratios remain the same 
in the intermediate polar V426 Oph, regardless of the isochoric/isobaric nature of the cooling-flow. \citet{2012A&A...538A..82R} also used an isochoric cooling-flow model for modeling plasma with temperatures hotter than 10$^{4}$ K. In addition, part of the geometry in IPs is controlled by moderate magnetic fields
\citep[][]{2011AcPol..51f..13G}.  An isochoric cooling-flow \textsc{Cloudy} model was, therefore, the simplest assumption for the application on V1223 Sgr.} The sample {\Cloudy} scripts used for generating the XSPEC compatible .fits files are given in the appendix.

Note that spectral signatures of X-ray reflection from the surface of the WD have been discussed in a number of studies \citep{1995MNRAS.272..749B, 1998MNRAS.293..222C, 2000MNRAS.315..307B, 2011PASJ...63S.739H, 2015ApJ...807L..30M, 2021MNRAS.504.3651H, 2021arXiv210705636I}. X-ray reflection primarily produce
two distinctive features:  a compton reflection hump at $\sim$ 10-30 keV, and iron fluorescent K$\alpha$ line at 6.4 keV. 
{\Cloudy} simulations include these features by default. 
The MEG spectrum covers the energy range 0.4 - 5.0 keV (31 - 2.5 \AA), where there are fluorescence lines from lighter elements. In case of V1223 Sgr, they are much fainter than the iron fluorescence line, thus negligible. This is not always true. For example, emission spectrum from the High Mass X-ray Binary Vela X-1 exhibits fluorescent Si and S K$\alpha$ lines \citep{2021A&A...648A.105A, 2021RNAAS...5..149C}.

\begin{figure*}
\gridline{\fig{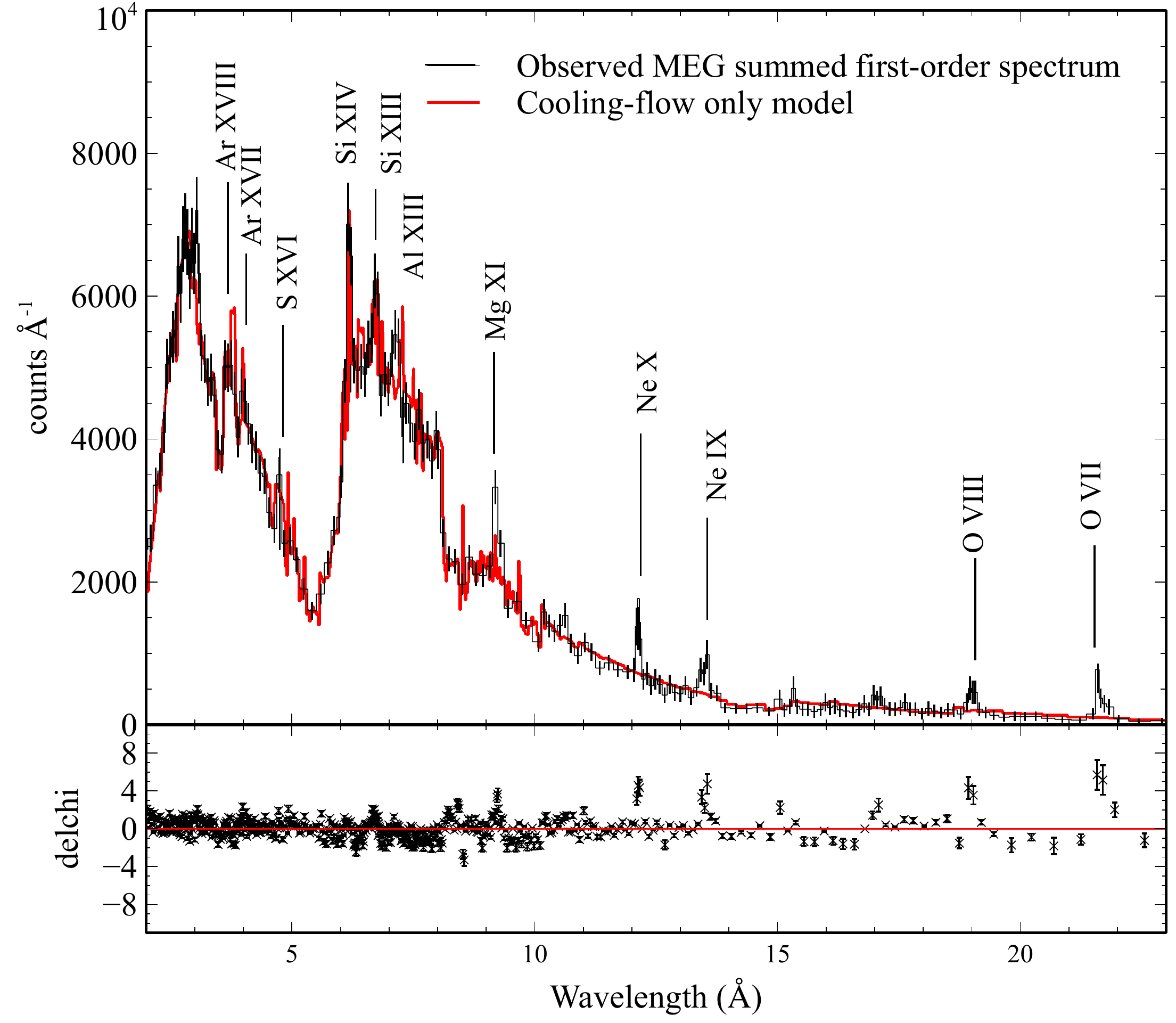}{0.5\textwidth}{(a)}
          \fig{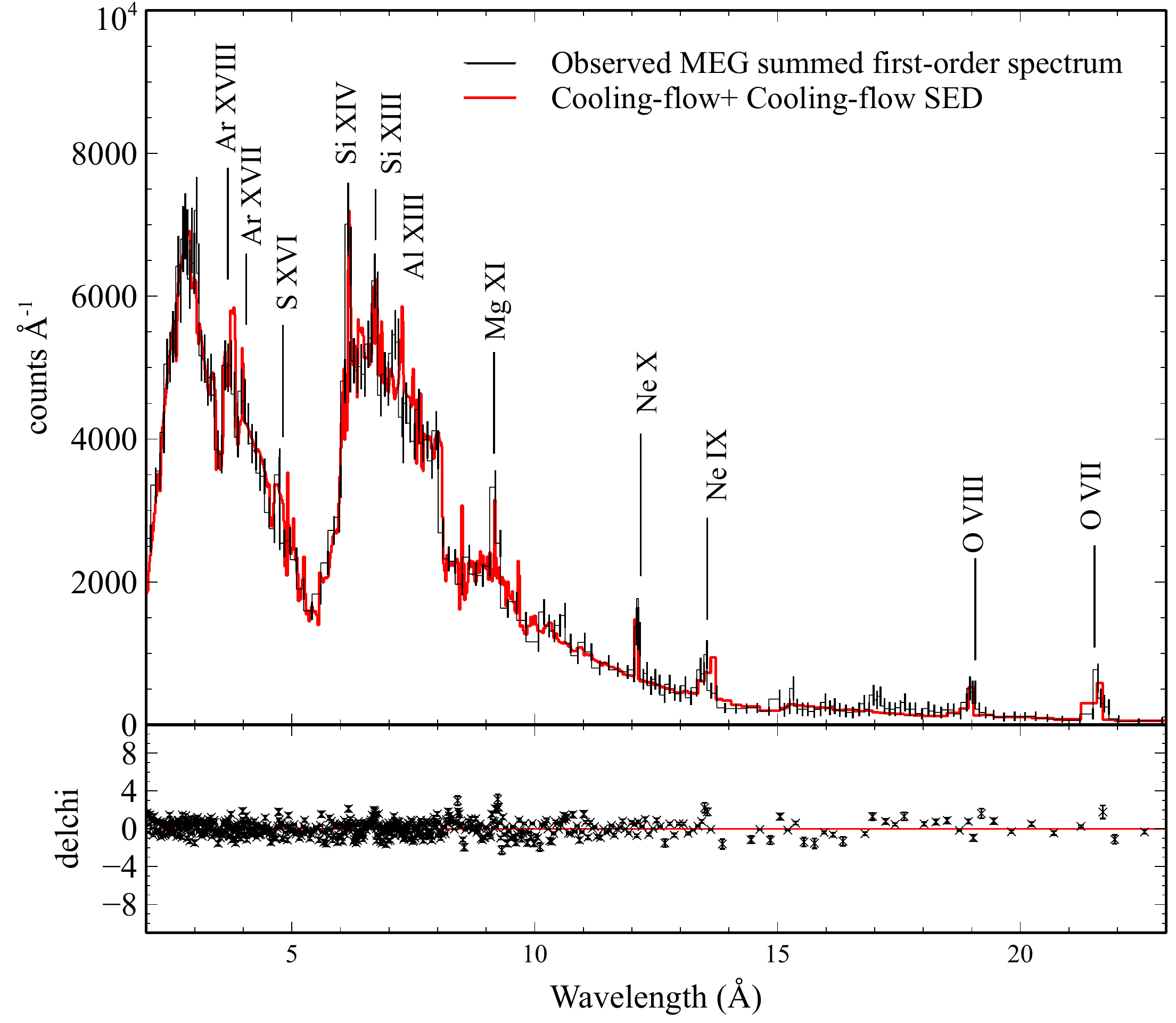}{0.5\textwidth}{(b)}
          }
\gridline{\fig{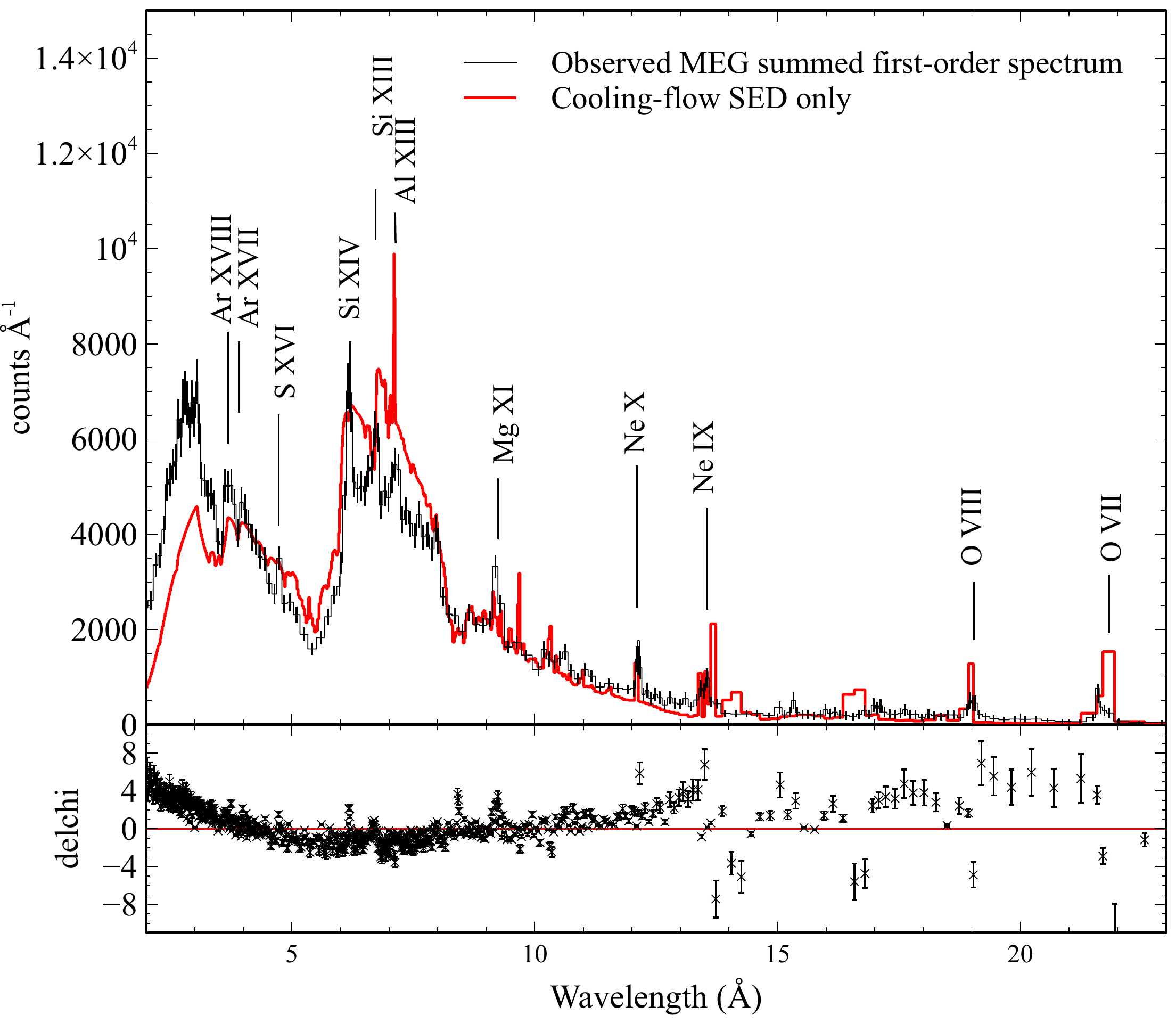}{0.5\textwidth}{(c)}
          \fig{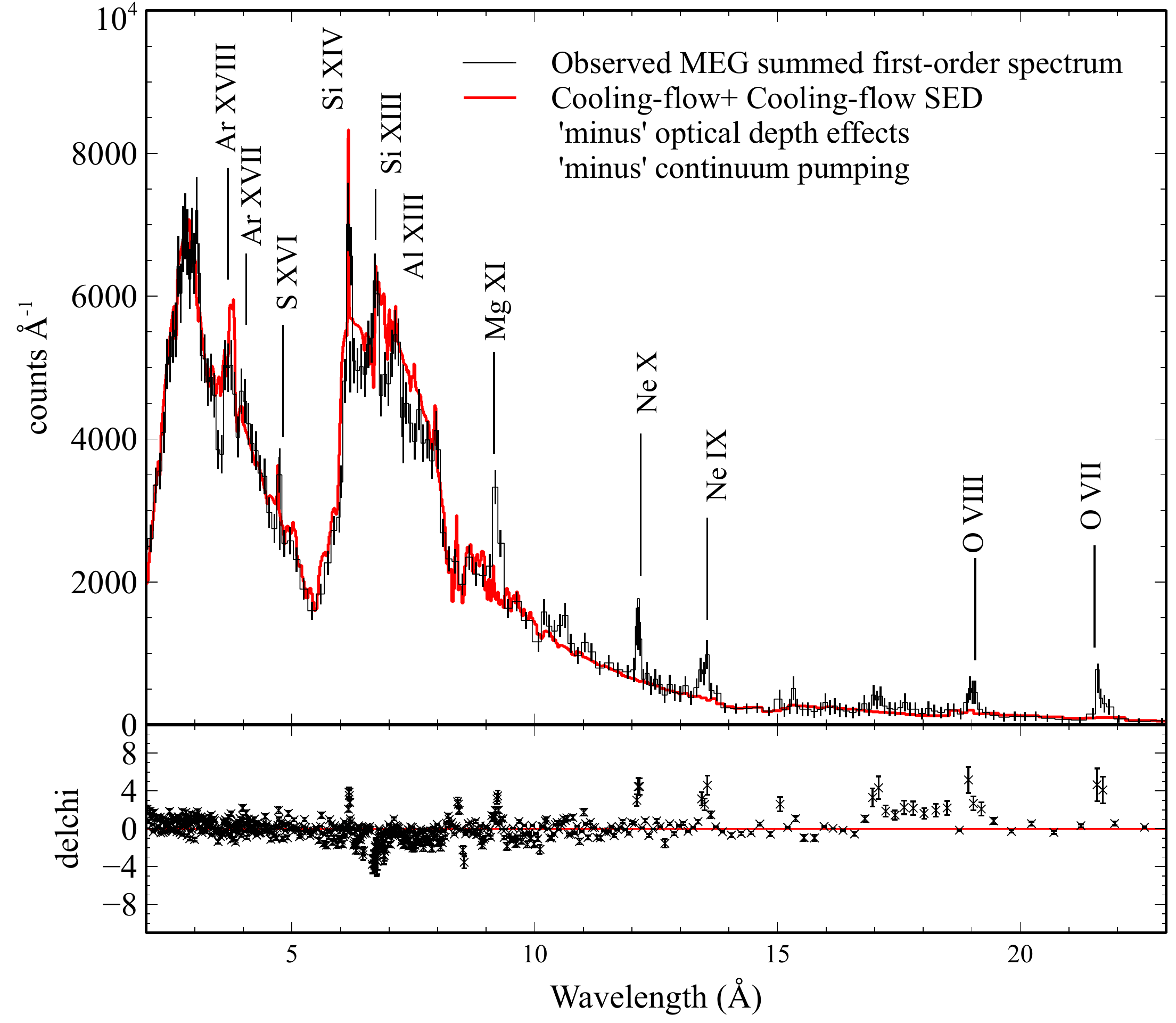}{0.5\textwidth}{(d)}
          }
\caption{On all four panels, black data points show the Chandra High Energy Transmission Grating MEG combined first-order spectrum. a) The solid red line shows our best-fit cooling-flow only \textsc{Cloudy} model. b) The solid red line shows
the best-fit cooling-flow + cooling-flow SED \textsc{Cloudy} model. c) The solid red line shows the cooling-flow SED only \textsc{Cloudy} model. d) The solid red line shows our best-fit cooling-flow + cooling-flow SED \textsc{Cloudy} model without considering the optical depth and continuum pumping effects discussed in Sections \ref{sec:linesuppression}, \ref{sec:enhancement}, and \ref{sec:softX-ray}.}
\label{fig:v1223}
\end{figure*}

The complete \textsc{Cloudy} models have been overplotted with the observed spectrum in Figure \ref{fig:v1223}. The red line represents our best-fit \textsc{Cloudy} model, and the black data points represent the observed spectrum. 
The top-left, top-right, bottom-left, and bottom-right panels show the best-fit \textsc{Cloudy} models using cooling-flow  only, cooling-flow + photoionized,
photoionized only, and cooling-flow + photoionized without considering the optical depth and continuum pumping.

(a) The cooling-flow only model fits the continuum well and contributes to the formation of  Al~XIII, Si~XIII, Si~XIV, S~XVI, Ar~XVII, and  Ar~XVIII lines. 
O~VII, O~VIII, Ne~IX,  Ne~X, and Mg~XI lines could not be fit with the cooling-flow model, which favors the findings of \citet{2021arXiv210705636I}.

%The green line shows the component of the complete model coming from photoionization, and the blue line shows the component coming from cooling-flow. 
(b) The cooling-flow + photoionized model produces a significantly better fit for O VII, O VIII, Ne IX,  Ne X, and Mg XI lines than the cooling-flow only model. The ionization fraction for O$^{7+}$, Ne$^{8+}$,  Ne$^{9+}$, and Mg$^{10+}$ peaks at temperatures greater  than 2 $\times$10$^{6}$ K, where we artificially stopped the cooling-flow \textsc{Cloudy} model. It is, therefore safe to infer that emission from O~VIII, Ne~IX,  Ne~X, and Mg~XI lines mostly comes from  photoionization. As the ionization fraction for O$^{6+}$ peaks at $\sim$ 8 $\times$10$^{5}$ K, we could not check if the cooling-flow model contributes to the 
O~VII line emission and to what extent. Future papers will address the numerical instability issue of the cooling-flow model at lower temperatures.

(c) The pure photoionized model produces a poor fit to the spectrum in the lower wavelength region with missing Si ~XIII, Si~XIV, S~XVI, Ar~XVII, and Ar~XVIII lines. 

(d) The cooling-flow + photoionized model with the optical depth effects and continuum pumping switched off produces 
a considerably worse fit to the observed spectrum than that including these atomic processes, especially the fit to the continuum
and emission lines from O~VII, O~VIII, Ne~IX,  Ne~X, Mg~XI, and Si~XIV. This shows the importance of taking into account the effects of optical depth and continuum excitation.

Table \ref{t:xspec}
lists \textsc{Cloudy} model parameters fitting the observed spectrum for (a), (b), (c), and (d) at a 90\% confidence interval. The best fit is obtained for model (b), with all the other fits being substantially worse.

%\begin{figure*}
%\gridline{\fig{v1223sgr_coolingflow_1.pdf}{0.5\textwidth}{(a)}
%          \fig{v1223sgr_coolingflow_sed_1.pdf}{0.5\textwidth}{(b)}
%          }
%\gridline{\fig{v1223sgr_sedonly_1.pdf}{0.5\textwidth}{(c)}
%          \fig{v1223sgr_Noopt_1.pdf}{0.5\textwidth}{(d)}
%          }
%\caption{On all four panels, black data points show the Chandra High Energy Transmission Grating MEG combined first-order spectrum. a) The solid red line shows %our best-fit cooling-flow only \textsc{Cloudy} model. b) The solid red line shows
%the best-fit cooling-flow + cooling-flow SED \textsc{Cloudy} model. c) The solid red line shows the cooling-flow SED only \textsc{Cloudy} model. d) The solid red %line shows our best-fit cooling-flow + cooling-flow SED \textsc{Cloudy} model without considering the optical depth and continuum pumping effects discussed in %Sections \ref{sec:linesuppression}, \ref{sec:enhancement}, and \ref{sec:softX-ray}.}
%\label{fig:v1223}
%\end{figure*}

\begin{table*}
\centering{
 % \captionsetup{}
         
\caption{\label{t:xspec}Best-fit parameters for: (a) cooling-flow only; (b) cooling-flow + cooling-flow SED; (c) cooling-flow SED only;  and (d) cooling-flow + cooling-flow SED model excluding continuum pumping and optical depth effects fitting the summed first-order MEG spectrum of V1223 Sgr.}
  }
\begin{tabular}{lrrrr}
\hline
\multicolumn{4}{c}{\hspace{200pt}Best-fit values} \\
\cline{2 -5}
Parameter & (a)  &  (b) & (c)  &  (d) \\

%w/o optical depth effects\\
\hline
%log $\alpha$ & - & -0.24$^{+0.01}_{-0.02}$  & -0.17$^{+0.02}_{-0.02}$ & 1.61 \\
log $\xi$ & - & 1.664 $^{+0.009}_{-0.007}$ & 0.853$^{+0.003}_{-0.004}$ & 1.612$^{+0.006}_{-0.005}$\\
log N$_{H_{photo}}$ & -& 22.93  $^{+0.03}_{-0.04}$ & 22.93$^{+0.04}_{-0.03}$& 23.39$^{+0.03}_{-0.03}$\\
N$_{ photo}$ ($\times$ 10$^{-24}$ ph keV$^{-1}$ cm$^{-2}$) & - & 0.46$^{+0.04}_{-0.04}$  & 6.51$^{+0.02}_{-0.01}$ & 0.25$^{+0.03}_{-0.04}$\\
Metallicity & 0.40${\pm 0.03}$ & 0.40${\pm 0.04}$ & 0.44${\pm 0.03}$& 0.45${\pm 0.03}$\\
N$_{ coll}$ ($\times$ 10$^{-24}$  ph keV$^{-1}$ cm$^{-2}$) &  4.86$^{+0.30}_{-0.20}$&2.23$^{+0.41}_{-0.41}$& -& 6.34$^{+0.32}_{-0.20}$ \\
log N$_{H_{coll}}$ & 23.13 $^{+0.03}_{-0.02}$ & 23.11 $^{+0.05}_{-0.04}$ & - & 23.08$^{+0.03}_{-0.04}$\\
$\chi^{2}$& 876.80/629 & 639.63/626 &  3233.08/628 & 988.35/626\\
\hline
\end{tabular}
 \end{table*}

\section{Summary}

\begin{itemize}
    \item 
     The effects of optical depth have been discussed on soft X-ray emission lines. A cloud
     can be collisionally-ionized  or photoionized.
    For a collisionally-ionized cloud, the atomic physical processes that contribute to the suppression in soft X-ray lines are photoelectric absorption of line photons and electron scattering. 
    Photoelectric absorption is temperature dependent, electron scattering is not. The temperature dependence of 
    photoelectric absorption opacity is shown on Figure \ref{fig:opacity}. The line modification factor ({$f_{\rm mod}$}) quantifies the fraction of photons surviving the absorption and scattering (see equation \ref{eqn:frac0}).
    Figure \ref{fig:supp} shows the numerical equivalent of  {$f_{\rm mod}$} as a function of 
    hydrogen column density and temperature for C~VI Ly-$\alpha$, O~VII K-$\alpha$ (w), O~VIII Ly-$\alpha$, and Fe~XVII lines. Tables \ref{t:c1} and \ref{t:c2} list {$f_{\rm mod}$} for important soft X-ray lines at N$_{H}$ = 10$^{22}$ cm$^{-2}$, N$_{H}$ = 10$^{22.5}$ cm$^{-2}$, and N$_{H}$ = 10$^{23}$ cm$^{-2}$ and log T = 6.4 and 6.8. 
    
    For a collisionally-ionized cloud, 0 $\leq$ $f_{\rm mod}$ $\leq$ 1. As shown in the tables, $f_{\rm mod}$ is smaller for lower-Z elements like carbon, oxygen, and neon, but approaches 1 for higher-Z elements like magnesium, silicon, and sulfur. This implies that soft X-ray line intensities for the lower-Z elements can be significantly suppressed.

    \item
    For the photoionized cloud, lines can be enhanced or suppressed. Continuum pumping can strengthen the Lyman lines in H- and He-like ions. 
    Absorption and scattering can suppress the lines. 
    Whether a line will be suppressed or enhanced is jointly determined by these two factors. Unlike the collisionally-ionized cloud, $f_{\rm mod}$ can be greater or smaller than 1. Figure \ref{f:photoionized} shows the numerical equivalent of  {$f_{\rm mod}$}  as a function of 
    hydrogen column density and ionization parameter for an $\alpha$=-1 power-law SED for O~VII K-$\alpha$ (w), O~VIII Ly-$\alpha$, Ne~IX K-$\alpha$ (w), and  Ne~X Ly-$\alpha$. Tables \ref{t:p1} and \ref{t:p2} show 
    $f_{\rm mod}$ for important soft X-ray photoionized lines 
for N$_{H}$ = 10$^{21}$ cm$^{-2}$, N$_{H}$ = 10$^{22}$ cm$^{-2}$, and N$_{H}$ = 10$^{23}$ cm$^{-2}$ for log $\xi$ = 1, and 2.5.

    \item
    
A hybrid of collisional ionization and photoionization is also observed in many astronomical sources. One such example is illustrated in Section \ref{obs}. We modeled the MEG spectrum from V1223 Sgr, an intermediate polar, using the \textsc{Cloudy}/XSPEC interface.  This is the first application of this interface 
after the recent \textsc{Cloudy} developments for meeting the spectroscopic standards of the future microcalorimeter missions.

We used a combination of  \textsc{Cloudy}-simulated cooling-flow (collisionally-ionized plasma cooling from 4$\times$10$^{8}$ K to 2$\times$10$^{6}$ K) and photoionized models to fit the spectrum. The photoionized model represents the emission from gas in the pre-shock region that has been ionized by the
diffuse emission from the cooling-flow plasma. The lower limit in temperature for the cooling-flow model was set to be 2$\times$10$^{6}$ K to avoid the numerical instability issues as described in Section \ref{multi-temp}.  We found that the cooling-flow model dominates the shape of the continuum. Emissions from Al~XIII, Si~XIII, Si~XIV, S~XVI, Ar~XVII, and  Ar~XVIII lines  originate in the cooling-flow plasma. O~VII, O~VIII, Ne~IX,  Ne~X, and Mg~XI lines originate in the photoionized plasma. \citet{2021arXiv210705636I} fitted the same spectrum of V1223 Sgr and reported excess emission from O~VII, O~VIII,  Ne~IX,  Ne~X, and Mg~XI lines which could  not be described with their absorbed cooling flow model. We fit these lines simply by adding the photoionized component to the standard cooling-flow component. Table \ref{t:xspec} lists the best-fit parameters from our best-fit \textsc{Cloudy} model. The top-right panel of Figure \ref{fig:v1223} shows the best-fit \textsc{Cloudy} model overplotted with the observed spectrum.

On the contrary, the observed spectrum was ill-fitted using pure cooling-flow and pure photoionized \textsc{Cloudy} models, as shown in the top-left and bottom-left panels of Figure \ref{fig:v1223}. We also show the consequences of switching off atomic physical processes like optical depth effects and continuum pumping in the bottom-right panel of the figure. The solid red line shows the fit after switching off the atomic physical processes. The fit is considerably worse than our best-fit model shown on the top-right panel, including all atomic processes. This shows the importance of incorporating optical depth effects and continuum pumping in the spectral simulation codes.

\end{itemize}

\section*{appendix A}
\textsc{Cloudy} script for generating XSPEC compatible cooling-flow model spectrum (where T$_{1}$ and T$_{2}$ are input parameters for temperature ):\\

\hspace{-3mm}\texttt{coronal T$_{1}$ K init time \# Plasma starts cooling from  T$_{1}$ K}\\
\texttt{hden 13.92}\\
\texttt{no scattering intensity}\\
\texttt{stop column density 20 vary}\\
\texttt{grid 20 24 0.5}\\
\texttt{metal solar -1 log vary}\\
\texttt{grid -1 0 0.2} \\
\texttt{abundances GASS10}\\
\texttt{iterate 3000}\\
\texttt{no molecules}\\
\texttt{stop time when temperature falls below T$_{2}$ K \# Plasma stops cooling at T$_{2}$ K}\\
\texttt{stop temperature T$_{2}$ K}\\
\texttt{save xspec atable total continuum "cooling-flow.fits"}\\

\textsc{Cloudy} script for generating  XSPEC compatible photoionized model spectrum:

\begin{verbatim}

table sed "cool.sed"
xi -1 vary
grid 0.5 2.5 0.5
hden 13.92
no scattering intensity 
stop column density 20 vary
grid 20 24 0.5
metal solar -1 log vary
grid -1 0 0.2
abundances GASS10
save xspec atable total continuum "photoionized.fits"
\end{verbatim}

\section*{appendix B}
Before version
17.03,  \textsc{Cloudy} used Chianti atomic database  \citep{1997A&AS..125..149D, 2012ApJ...744...99L} as described in \citet{2015ApJ...807..118L} for Fe$^{16+}$  energy levels. With 17.03, we updated the masterlist to use the first 30 levels for Fe$^{16+}$ from the NIST atomic database \citep{2018APS..DMPM01004K} by default, which improved the wavelength for the Fe XVII 3C$^\dagger$ transition from 15.0130\AA~ to 15.0150\AA.

\section*{Acknowledgement}

We acknowledge support by NSF (1816537, 1910687), NASA (17-ATP17-0141, 19-ATP19-0188), and STScI (HST-AR-15018).
MC also acknowledges support from STScI (HST-AR-14556.001-A). SB acknowledges financial support from the Italian Space Agency (grant 2017-12-H.0) and from the PRIN MIUR project ‘Black Hole winds and the BaryonLife Cycle of Galaxies: the stone-guest at the galaxy evolution supper’, contract 2017-PH3WAT.

\bibliography{sample631}{}
\bibliographystyle{aasjournal}

\end{document}